\documentclass[twocolumn,showpacs,amsmath,amssymb,superscriptaddress]{revtex4}

\usepackage[T1]{fontenc}
\usepackage[cp1250]{inputenc}
\usepackage{graphicx}
\usepackage{dcolumn}
\usepackage{bm}
\usepackage{ulem}

\begin{document}

\title{Nonlinear Magneto-Optical Rotation with Amplitude-Modulated Light: AMOR}

\author{W.~Gawlik}
\affiliation{M.~Smoluchowski Institute of Physics, Jagiellonian University, Reymonta 4, 30-059 Kraków, Poland}
\author{L.~Krzemień}
\affiliation{M.~Smoluchowski Institute of Physics, Jagiellonian University, Reymonta 4, 30-059 Kraków, Poland}
\author{S.~Pustelny}
\affiliation{M.~Smoluchowski Institute of Physics, Jagiellonian University, Reymonta 4, 30-059 Kraków, Poland}
\author{D.~Sangla}
\altaffiliation[on leave from ]{Ecole Superieure d'Optique, Orsay, France.} \affiliation{M.~Smoluchowski Institute
of Physics, Jagiellonian University, Reymonta 4, 30-059 Kraków, Poland}
\author{J.~Zachorowski}
\affiliation{M.~Smoluchowski Institute of Physics, Jagiellonian University, Reymonta 4, 30-059 Kraków, Poland}
\author{M.~Graf}
\affiliation{Department of Physics, University of California at Berkeley, Berkeley, CA 94720-7300, USA}
\author{A.O.~Sushkov}
\affiliation{Department of Physics, University of California at Berkeley, Berkeley, CA 94720-7300, USA}
\author{D.~Budker}
\affiliation{Department of Physics, University of California at Berkeley, Berkeley, CA 94720-7300, USA}
\affiliation{Nuclear Science Division, Lawrence Berkeley National Laboratory, Berkeley CA 94720, USA}

\date{\today}

\begin{abstract}
A new technique of nonlinear magneto-optical rotation with amplitude modulated light is developed.
The technique is an alternative to its counterpart with frequency modulated light and can be applied to sensitive
measurements of magnetic
fields ranging from microgauss to the Earth-field level. The rotation signals exhibit nontrivial features like
narrowed non-Lorentzian lineshapes and multi-component resonances.
\end{abstract}

\maketitle

We report on a study of resonant nonlinear magneto-optical rotation in rubidium vapor with amplitude modulated light. The
Amplitude Modulated magneto-Optical Rotation (AMOR) has been measured by lock-in detection in modulated
transmitted light as a function of  magnetic field, light intensity, size of the absorption cells, their isotopic
content, and modulation waveforms.

First applications of the amplitude modulation (AM) of light intensity to atomic and molecular spectroscopy go
back to the synchronous optical pumping studies of Bell and Bloom \cite{Bell61} and Series \cite{Ser64}. Since
then the method has been applied in many situations, see Ref. \cite{Alex05} for a review.

In the present work we concentrated on using AM light and optical pumping synchronous with Larmor precession for
studies of the nonlinear magneto-optical rotation \cite{Bud02RMP} and on establishing an alternative to the frequency-modulated-light nonlinear magneto-optical rotation (FM NMOR) \cite{Bud02PRA}. In both methods
ultra-narrow resonances at zero magnetic field are replicated at higher fields. Thus they are suitable as
magnetometry techniques for the fields ranging from the microgauss level to the Earth field with sensitivity reaching $\sim10^{-12}\:G/\sqrt{Hz}$. While the FM NMOR method is straightforward to implement and sensitive to highly nonlinear magneto-optical effects \cite{Yas02}, it is prone to distortions by a possible spurious AM modulation of laser light \cite{Gil01} and the AC Stark-effect shifts. These shifts
caused by intense light occur in FM NMOR where the light frequency is tuned to the side of a resonance line,
but can be minimized in AMOR which uses resonant excitation.
AMOR technique can be used when it is impossible or difficult to change the light-source frequency; AM also offers additional freedom in optimizing the modulation waveform for better control of the atomic dynamics and observed signals.

\begin{figure}[b]
  \centering
  \includegraphics[width=7 cm]{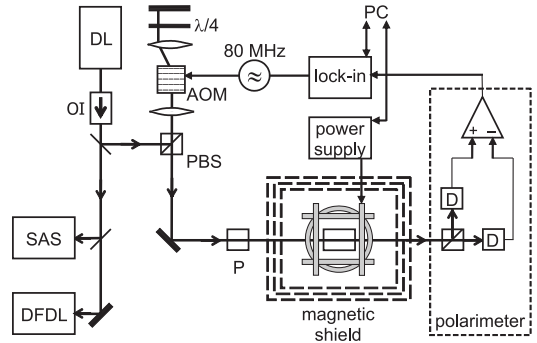}\\
  \caption{Experimental set-up. DL stands for external-cavity diode laser, OI --- optical isolator, SAS --- saturated absorption frequency reference, DFDL --- Doppler-free dichroic lock, D --- photodiodes, P --- crystal polarizer, PBS --- polarizing beam splitter, $\lambda$/4 --- quarter-wave plate, and PC is a computer controlling the experiment.}\label{setup}
\end{figure}

The layout of the experimental setup is presented in Fig. \ref{setup}. Experiments were performed with rubidium
atoms in two different glass cells: one of 5-cm length and 2.5-cm diameter containing natural isotopic mixture of
$^{85}$Rb and $^{87}$Rb, and the second of 1.5-cm length and 1.8-cm diameter containing $^{87}$Rb. Both cells
contained 3~torr of Ne as a buffer gas.

The cells were placed within a three-layer $\mu$-metal magnetic shield and were surrounded by several coils used
to produce magnetic field $B$ along the laser beam and to compensate transverse fields.

As the light source we used an external-cavity diode laser locked to the center of the $F=2\rightarrow F'=2$ hfs
component of the $D_1$ transition $5\:^2S_{1/2}\rightarrow 5\:^2P_{1/2}$ (795~nm). The main part of the beam was
transmitted in a double pass through an acousto-optical modulator (AOM) driven by an 80~MHz radio-frequency signal
whose amplitude was modulated with various waveforms of frequency $\Omega_m$. In this way the laser light was
frequency-shifted by 160~MHz and its intensity $I$ was modulated at $\Omega_m/2\pi$ ranging from $20$ to $100$~kHz
and with different modulation depths $m=(I_{max} - I_{min})/ I_{max}$. The results presented below were obtained
at $\Omega_m/2\pi=30$~kHz. The linearly polarized beam of 2-mm diameter traversed the cell placed in a magnetic
shield and kept at 20°C. The nonlinear magneto-optical rotation  was analyzed by a polarimeter consisting of a
Glan prism and two photodiodes detecting two orthogonally polarized components of the transmitted light. The polarimeter was balanced by
rotating it at 45° relative to the initial beam polarization. The difference signal was analyzed by a lock-in
detector whose in-phase and quadrature outputs at the first harmonic of $\Omega_m$, were stored in a PC as
functions of the magnetic field.

\begin{figure}
  \centering
  \includegraphics[width=7cm]{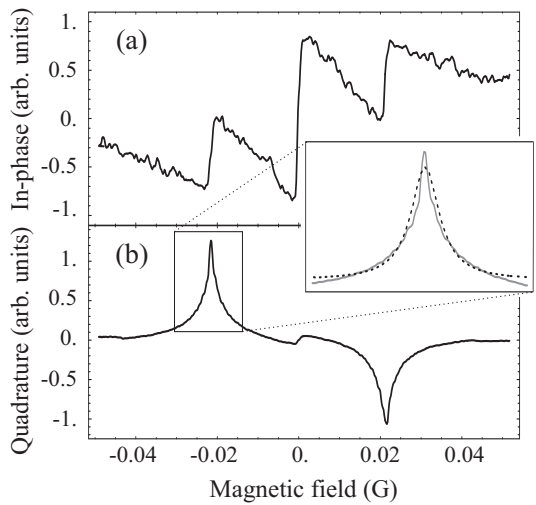}\\
  \caption{Short-cell, $^{87}$Rb signals taken with 1.5-cm long cell and sine-wave modulation with $m=25\%$ and light power of 10~$\mu$W. In-phase component (a), quadrature component (b). The maximum rotation angle is about 0.25 mrad. Note that both components are shown on the same scale but have different S/N ratios. The inset shows the Lorentzian fit to the sideband resonance as a dashed line.} \label{sine_modul}
\end{figure}

The AMOR signals recorded with a sine-wave AM as a function of magnetic field $B$ (Fig.~\ref{sine_modul}) have a
characteristic form consisting of a central resonance at $B=0$ and two side resonances (sidebands) appearing for
the resonance condition $\Omega_L= \Omega_m/2$, where $\Omega_L=g\mu_BB/\hbar$ is the Larmor frequency, $\mu_B$ being the Bohr magneton and $g$ the Land\'{e} factor. The signal form is essentially identical with that seen with FM NMOR \cite{Bud02PRA}. New features appear with square-wave modulation and are described in detail below.

The widths of the resonances depend on several factors: the light power, Rb diffusion in buffer gas, and the
magnetic-field inhomogeneity. The narrowest widths (FWHM) obtained with the present non-optimized setup are about
2.1~mG for the short cell and about 2.6~mG for the long cell. The resonances seen with both cells differ not only in
their widths but also in the lineshapes. The short-cell resonances have non-Lorentzian shapes (see the inset in
Fig.~\ref{sine_modul}), similar to those studied recently in electromagnetically induced transparency by Xiao et al.
\cite{Xiao05} and attributed by these authors to diffusion-induced Ramsey narrowing. Indeed, the resonance widths
determined by diffusion time of Rb atoms in Ne gas of 3~torr pressure through a 2-mm wide beam
should be about 5 times larger than what we measure. On other hand, the long-cell resonances exhibit closer to
Lorentzian lineshapes which we explain by stronger effects of the magnetic field inhomogeneities that accumulate
over the longer cell length and overwhelm the diffusion-induced Ramsey narrowing.

\begin{figure}
  \centering
  \includegraphics[width=7cm]{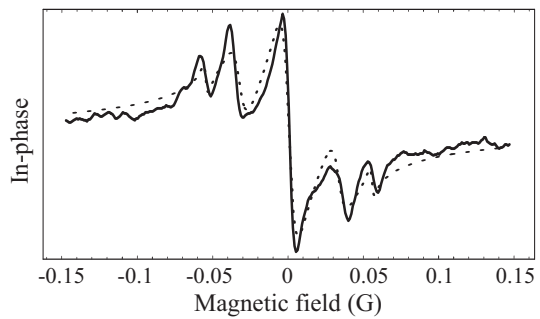}\\
  \caption{AMOR resonances with the 5-cm cell containing natural rubidium obtained with light power of 15~$\mu$W. The sidebands occurring at higher magnetic fields are associated with $^{85}$Rb while the ones that are closer to $B=0$ are due to $^{87}$Rb. Dashed line depicts a fit with a set of dispersive Lorentzians.\
}\label{double}
\end{figure}

Using the 5-cm cell with natural rubidium and maintaining the light power below 60~$\mu$W to reduce power broadening, we recorded resolved sideband resonances associated with the two Rb isotopes. Since
the Land\'{e} factors and Larmor frequencies for the two isotopes are different [g($^{85}$Rb)$\approx$1/3,
g($^{87}$Rb)$\approx$1/2], the resonances occur at different magnetic fields which gives rise to double sidebands, as
shown in Fig. \ref{double}. The sideband positions yield the ratios which deviate by about 10\% from the $2/3$
ratio of the Land\'{e} factors of the two isotopes
and weakly depend on the light power. The origin of this discrepancy is currently under investigation.

\begin{figure*}
  \centering
  \includegraphics[width=17.5cm]{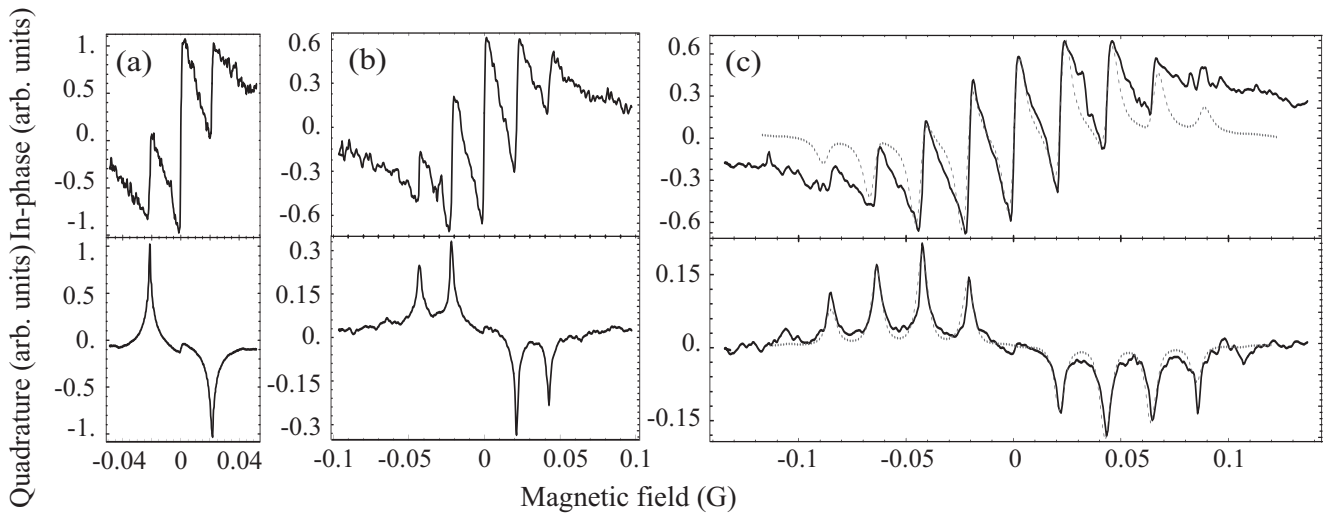}\\
  \caption{AMOR resonances recorded with the 1.5-cm $^{87}$Rb cell, square-wave 100\% modulation of 0.5 duty cycle, and 10~$\mu$W average power (a), 0.35 duty cycle and 6~$\mu$W average light power (b), and 0.2 duty cycle and 5~$\mu$W average light power (c). Upper row shows in-phase signals, lower row shows quadrature signals. The broken lines show theoretical fits.}\label{square_modul}
\end{figure*}

The square-wave modulation leads to occurrence of additional ``harmonic'' resonances at magnetic fields
corresponding to $\Omega_L$=$n\Omega_m/2$ with $n$ being an integer >1. These additional resonances arise due to the presence in the modulation waveform of Fourier components at multiples of the modulation frequency. While
single sidebands are visible for modulation with duty cycle 0.5 within our range of the magnetic field scan
(Fig.~\ref{square_modul}a), for duty cycle 0.35 the first and second harmonics are clearly seen with comparable
amplitudes (Fig. \ref{square_modul}b). When the duty cycle of the square-wave modulation decreases, i.e. the light
pulses are shorter, the number of harmonics increases. Fig.~\ref{square_modul}c shows this effect for
a duty cycle of 0.2 when up to five harmonics are well visible under conditions similar to those of
Figs.~\ref{square_modul}a and \ref{square_modul}b. The experimental recordings are compared to a theory based on the  model of Ref. \cite{Kan}. Time-dependent solutions of the model are multiplied by the sine and cosine reference signals and integrated to simulate the lock-in signals. The calculated signals (broken line) reproduce most salient features of the experimental ones except for a residual resonance seen at $B=0$ in
the quadrature component. The amplitude of this residual resonance increases with light intensity, thus
we attribute it to the alignment-to-orientation conversion \cite{Bud00, Bala}. It is noteworthy that, unlike
in FM NMOR, the in-phase component is noisier than the quadrature one.

Our study demonstrates the applicability of the AMOR technique to studies of magneto-optical phenomena which
are highly sensitive to ground-state level shifts, for example, due to a magnetic field.

The AMOR method has the same possible applications as the FM NMOR. In particular, it should be useful in
Earth-field and space magnetometry, in applications with miniature cells and in NMR/MRI \cite{Yas04}.
The use of low duty-cycle square-wave modulation (train of short pulses) allows studies of the dynamics of the induced atomic observables and results in appearance of many equally spaced sideband resonances.  These resonances, occurring at harmonics of Larmor frequency, reach high magnetic fields even with low-frequency of light amplitude modulation which significantly reduces experimental requirements. The resonances have maximum amplitudes at resonant frequency of laser light which reduces possible distortions caused by AC-Stark effect. These features make the AMOR technique a useful method of nonlinear magneto-optics and an attractive possible alternative to FM NMOR.

The authors thank M.~Auzinsh, M.~Ledbetter, and I.~Novikova for their valuable comments on the manuscript.  This work has been supported by Polish grant KBN 3T11B/07926, and by the Jagiellonian University, ONR MURI, and an NSF US-Poland collaboration grant. Participation of the US students (M.~G. and A.~S.) has been sponsored by the NSF Global Scientists program.

\end{document}